\documentclass[a4paper,fleqn,usenatbib]{mnras}

\usepackage{graphicx}	
\usepackage{amsmath}	
\usepackage{amssymb}	
\usepackage{lscape}
\usepackage{textcomp}
\usepackage{multirow}
\usepackage{times}
\usepackage{booktabs}
\usepackage{deluxetable}

\def\ltsima{$\; \buildrel < \over \sim \;$}
\def\gtsima{$\; \buildrel > \over \sim \;$}
\def\lsim{\lower.5ex\hbox{\ltsima}}
\def\gsim{\lower.5ex\hbox{\gtsima}}
\def\lapp{\ifmmode\stackrel{<}{_{\sim}}\else$\stackrel{<}{_{\sim}}$\fi}
\def\gapp{\ifmmode\stackrel{>}{_{\sim}}\else$\stackrel{<}{_{\sim}}$\fi}

\def\msol{\,\mathrm{M}_\odot}

\title[]{The chemical composition of the low-mass Galactic globular cluster NGC~6362\footnote{ 
Based on FLAMES observations performed at the European Southern Observatory, 
proposal numbers 093.D-0618(A) and 097.D-0325(A).}}

\author[D.Massari et al.]{
D. Massari,$^{2,3,4}$\thanks{E-mail: massari@astro.rug.nl}
A. Mucciarelli,$^{5,4}$
E. Dalessandro,$^{4,5}$
M. Bellazzini,$^{4}$
S. Cassisi,$^{6}$
\newauthor G. Fiorentino,$^{4}$
R.A. Ibata,$^{7}$
C. Lardo,$^{8}$
M. Salaris$^{8}$
\\
$^{2}$University of Leiden, Leiden Observatory, NL$-$2300 RA Leiden, Netherlands\\
$^{3}$University of Groningen, Kapteyn Astronomical Institute, NL$-$9747 AD Groningen, Netherlands\\
$^{4}$INAF-Osservatorio Astronomico di Bologna, via Ranzani 1, I$-$40127, Bologna, Italy\\
$^{5}$Dipartimento di Fisica e Astronomia, Universit\`a degli Studi di Bologna, v.le Berti Pichat 6/2, I$-$40127 Bologna, Italy\\
$^{6}$INAF-Osservatorio Astronomico di Teramo, Via M. Maggini, I$-$64100 Teramo, Italy\\
$^{7}$Observatoire astronomique de Strasbourg, Université de Strasbourg, CNRS, UMR 7550, 11 rue de l’Université,
F$-$67000 Strasbourg, France\\
$^{8}$Astrophysics Research Institute, Liverpool John Moores University, IC2, Liverpool Science Park, 146 Brownlow Hill, Liverpool L3 5RF, UK\\
}

\date{Accepted XXX. Received YYY; in original form ZZZ}

\pubyear{2017}

\begin{document}
\label{firstpage}
\pagerange{\pageref{firstpage}--\pageref{lastpage}}
\maketitle

\begin{abstract}
We present chemical abundances for 17 elements
in a sample of 11 red giant branch stars in NGC~6362 from UVES spectra. 
NGC~6362 is one of the least massive globulars where multiple populations 
have been detected, yet its detailed chemical composition has not been investigated so far.
NGC~6362 turns out to be a metal-intermediate ([Fe/H]=--1.07$\pm$0.01 dex) cluster, 
with its $\alpha$- and Fe-peak elements content compatible with that observed 
in clusters with similar metallicity. It also displays an enhancement in 
its s-process element abundances. 
Among the light elements involved in the multiple populations phenomenon, 
only [Na/Fe] shows star-to-star variations, while [Al/Fe] and [Mg/Fe] 
do not show any evidence for abundance spreads.
A differential comparison with M4, a globular cluster with similar mass and metallicity, 
reveals that the two clusters  share the same chemical composition.
This finding suggests that NGC~6362 is indeed a regular cluster, formed from gas 
that has experienced the same chemical enrichment of other clusters 
with similar metallicity.
\end{abstract}

\begin{keywords}
globular clusters: individual (NGC~6362) - stars: abundances - techniques: spectroscopic
\end{keywords}



\section{INTRODUCTION}

With ages of the order of $12$-$13$ Gyr, Globular clusters (GCs) are thought to be among the first stellar systems formed at 
early epochs in the Local Group. 
Thanks to the general homogeneity in terms of age and chemical composition of their stars,
GCs have been used for decades as ideal tracers of the chemistry 
of their environments, allowing us to reconstruct the chemical enrichment history and the age-metallicity 
relation of their host galaxies.

The current picture of GC formation and evolution moves away from the traditional paradigm 
of GCs as simple stellar population \citep[see for instance the seminal paper by][]{renzini86}, according to which 
all the stars in a GC share the same initial chemical abundances for all the chemical elements.
In fact, the recent discoveries of multiple sequences in GC 
colour-magnitude diagrams (CMDs, see e.g. \citealt{piotto09, piotto15}) and of star-to-star variations in the 
chemical abundances of some light elements  \citep[like C, N, Na, O, Mg, Al, see e.g.][and references therein]{gratton12} 
demonstrate that such systems are indeed much more complex.
Multiple populations (MPs) are ubiquitous in all GCs studied so far, 
both in the Galactic (see for example \citealt{carretta09, gratton12, piotto15}) and extra-Galactic (\citealt{muccia09, dalex16}) environment. 
Since this chemical pattern has been observed in stars at all the evolutionary stages (\citealt{gratton04, milone12a}), 
it cannot be explained in terms of internal mixing but it should have been imprinted in stars at formation.
Several scenarios interpret the photometric and spectroscopic evidence in terms of different generations of stars,
with a first generation polluting the gas out of which second 
generation of stars formed with the products of their internal evolution. 
A number of candidate polluters have been proposed (see \citealt{renzini15} for a comprehensive review); namely 
asymptotic giant branch stars (\citealt{dercole08}), fast-rotating massive stars (\citealt{decressin07}), 
massive binaries (\citealt{demink09}) and supermassive stars (\citealt{denissenkov}).
Alternative scenarios have been proposed as well, where low-mass stars accrete the polluted material during 
the pre-main sequence phase to give rise to a chemically peculiar population of stars that is coeval to that with pristine composition
(\citealt{bastian13}).

The abundance spreads typical of MPs involve only a few light elements. 
For most of the elements (in particular $\alpha$-elements like Si, Ca and Ti, iron-peak and 
neutron-capture elements) the stars in a typical GC exhibit a remarkable level of internal homogeneity, 
thus suggesting that the abundances of these elements in GCs can be used to trace 
the chemical composition of the gas from which the cluster formed. 
Therefore, the investigation through high-resolution spectroscopy of the chemistry of poorly (or not yet) 
studied clusters is crucial to understand the chemical evolution of the Galactic Halo 
\citep[see e.g.][]{pritzl05,apogee15}
and to identify, through the chemical tagging, clusters that likely originated in extragalactic environments
\citep{muccia13,villanova13,munoz13,carretta14,marino15}.

In this paper, we present the first detailed chemical study of NGC~6362, an intermediate metallicity clusters which belongs 
to the low-mass tail of the GC mass distribution, with a mass of only $5\times10^{4}M_{\odot}$ \citep{dalex14}. 
We list cluster fundamental properties in Tab.~\ref{tab0}, along with other useful informations.
\citet{muccia16} firstly measured its iron content ([Fe/H]= --1.09 $\pm$ 0.01 dex) by analysing 160 giant stars observed 
with FLAMES at the Very Large Telescope. They also found a bimodal [Na/Fe] distribution, that makes NGC~6362 one of the least massive 
clusters where MPs have been detected.
The peculiar radial distribution of its stars, with both first and second populations
being completely mixed out to several half light radii, has been interpreted by \cite{dalex14} as the 
result of heavy mass-loss due to long-term dynamical evolution.
Such a claim is also supported by the quite shallow present-day mass function observed
for this cluster in \cite{paust10}. Thus, it would be extremely interesting to investigate  whether it also 
shows other exceptional chemical features or it follows the general trends
observed for standard GCs in the Milky Way with MPs.
To this end, we here present a detailed chemical analysis of  11 members in NGC~6362.

The paper is organized as follows. The dataset analysed in this work is described in Section \ref{data}, 
while the details of the chemical analysis are discussed in Section \ref{analysis}. 
We present the results of this study in Section \ref{res} and we finally discuss them in Section \ref{discuss}.

\section{OBSERVATIONS AND DATA REDUCTION}\label{data}

\begin{figure}
\includegraphics[width=\columnwidth]{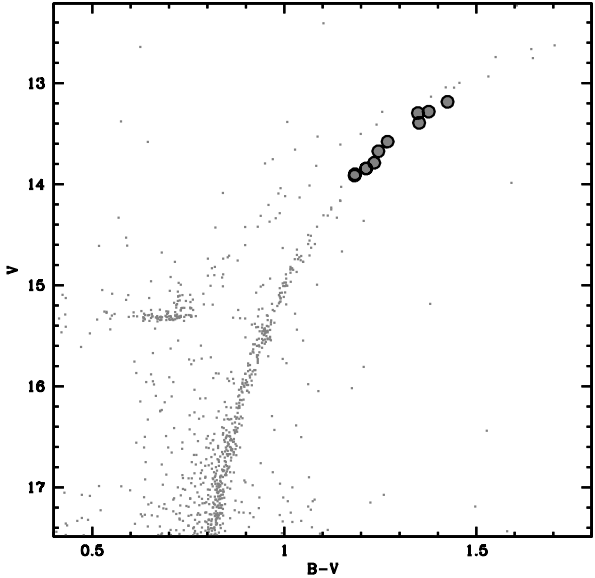}
\caption{\small Optical (V, B-V) CMD of NGC~6362 in the innermost 300\arcsec (the photometry comes from the catalogue described in
\citealt{dalex14}). The UVES targets are highlighted with black filled circles.}
\label{cmd_targ}
\end{figure}

The dataset analysed in this work was acquired under the programs 093.D-0618(A) and 097.D-0325(A), PI: Dalessandro,  
with the FLAMES spectrograph (\citealt{pasquini}) at the ESO Very Large Telescope.
We used the combined MEDUSA+UVES configuration, which allowed the simultaneous allocation of
eight UVES high-resolution fibres and 132 MEDUSA mid-resolution fibres per exposure.
While the stars observed in the MEDUSA mode have been discussed in \cite{muccia16},
in this work we focus on the eleven stars observed with the UVES 580 Red Arm spectral configuration,
with a resolution of R$\sim$45000 in the spectral range $\sim4800-6800$ \AA{}.
Targets have been selected from the Wide Field Imager (WFI) photometric catalogue presented in
\cite{dalex14}. For this study, only red giant branch (RGB) stars brighter than V$<14$ mag have been targeted (see Figure \ref{cmd_targ}).
Also, to avoid contamination from neighbours, only stars with no close (i.e. within 2\arcsec) sources brighter than V$<$ V$_{target}+1$ 
 have been selected.
Four exposures of 45 min for each target have been secured. Two UVES fibres have been used to sample the sky
background, thus allowing a proper sky subtraction for each individual exposure.

\begin{deluxetable}{cc}
\tablewidth{0pc}
\tablecolumns{2}
\tablecaption{Fundamental properties of NGC~6362.}
\tablehead{\colhead{~~~~~~~~~~~~~~~~~~~~~~~~~NGC~6362}}
\startdata
        Right Ascension (h:m:s) & $17$:$31$:$54.99$          \\
        Declination ($^\circ$:\arcmin:\arcsec) & -$67$:$02$:$54.0$  \\
        Distance (kpc) & $7.6$  \\
        r$_{h}$ (\arcmin) & $2.5$  \\
        Mass ($\msol$) & $5.3\times10^{4}$  \\
        $[$Fe/H$]$ (dex) & $-1.07$ \\
\enddata
\tablecomments{\small Position (\citealt{goldsbury10}), distance (\citealt{harris}), half-light radius and mass (\citealt{dalex14}) and metallicity (this work) of NGC~6362 }
\label{tab0}
\end{deluxetable}

To reduce the acquired data, we used the last version of the FLAMES-UVES Common Pipeline Libraries 
based ESO pipeline\footnote{http://www.eso.org/sci/software/pipelines/},
which includes bias-subtraction, flat-field correction, wavelength calibration with a standard
Th-Ar lamp, extraction of one-dimensional spectra and order merging.
The accuracy of the dispersion solution has been checked by comparing the observed position of several 
sky emission lines with their rest-frame position as reported in the sky lines atlas by \cite{oster96}. 
No significant wavelength shifts have been found.
Once extracted, all the individual (sky-subtracted) exposures of each target have been brought
to the same reference by correcting for the corresponding heliocentric radial velocity (see Sect.\ref{vrad}
for the details), and finally combined together to obtain a median spectrum. The final reduced spectra have signal-to-noise ratio (SNR) larger than $\sim$30 at any wavelength.

\subsection{Radial velocities}\label{vrad}

Radial velocities (v$_{rad}$) have been measured using the wrapper 4DAO\footnote{4DAO is freely distributed at the website 
{\tt http://www.cosmic-lab.eu/4dao/4dao.php.}} (\citealt{4dao}), which allows to run DAOSPEC \citep{daospec} for
large sets of spectra, tuning automatically its main input parameters.
For all the eleven targets, we treated the two UVES chips of each single exposure independently. For the lower (L) chip, v$_{rad}$ have been
computed by using $\sim 160$ absorption lines, while for the upper (U) chip the lines used were $\sim190$.
For each target we obtained a remarkably good agreement between the measurements of the two independent chips, 
finding an average difference of only v$_{rad,U}$-v$_{rad,L}=+0.06$ km$\,$s$^{-1}$ ($\sigma=0.18$ km$\,$s$^{-1}$). 
Moreover, we did not find any significant difference among the v$_{rad}$ measured for the four exposures
of each target. This indicates that none of the observed targets are binary systems.
We then computed the final values of v$_{rad}$ as the average value of the eight single exposure measurements 
(two chips for each of the four exposures) and we adopted the dispersion around the mean divided by $\sqrt(8)$ as uncertainty.
Heliocentric velocities and related errors are listed in Table \ref{tab1}.

We measured an average v$_{rad}$ of v$_{rad}=-15.03$  km$\,$s$^{-1}$ ($\sigma=2.07$  km$\,$s$^{-1}$), which is 
in agreement with the value derived by \cite{muccia16} (v$_{rad}=-13.8$ km$\,$s$^{-1}$, $\sigma=2.7$ km$\,$s$^{-1}$).
According to their v$_{rad}$ distribution all the targeted stars are members, 
having v$_{rad}$ values that lie well within 2$\sigma$ from the systemic velocity.

\section{CHEMICAL ANALYSIS}\label{analysis}

\subsection{Atmospheric parameters}\label{atm}

Effective temperatures (T$_{{\rm eff}}$) and surface gravities ($\log$ $g$) for target stars 
have been derived from their the B and V magnitudes, in the same way as described in 
in \cite{muccia16}. We correct magnitudes and colours for differential reddening using the procedure outlined in \citet{massari12} \citep[see also][]{milone12}. Differential reddening corrections across the whole WFI field of view range 
from $\delta[E(B-V)]=-0.03$ mag to +0.03 mag around the adopted average colour excess E(B-V)$=0.09$ mag (\citealt{reed88}).

Errors on the parameters affecting the determination of the
absolute colour for the analysed targets, i.e photometric errors or uncertainty on the absolute and differential reddening ($\sigma_{{\rm[E(B-V)]}}$ and  $\sigma_{{\rm \delta[E(B-V)]}}$, respectively) could potentially affect our 
T$_{{\rm eff}}$ estimates. Thus, to evaluate the uncertainties on $T_{{\rm eff}}$
we re-determined temperatures assuming typical errors on colors and extinction of $\sigma_{{\rm B,V}}=0.01$ mag, $\sigma_{{\rm[E(B-V)]}}=0.04$ mag and $\sigma_{{\rm \delta[E(B-V)]}}=0.02$ mag, respectively.
We underline that the quoted values are conservative upper limits, since the nominal photometric errors
of such bright and well exposed stars are $<0.01$ mag, while the errors on the absolute and differential 
reddening estimates correspond to about the 50\% of their value (see \citealt{dalex14}). 
Finally, we measure a typical uncertainty on T$_{{\rm eff}}$ of $\sim90$ K.

Stars located in the brightest portion of the RGB, especially those approaching the RGB tip, can be significantly be affected by
non-LTE effects that spuriously decrease the iron abundances from FeI lines leaving
those from FeII lines unaltered (see \citealt{ivans01,lapo14,m3201}).
This causes the spectroscopic derivation of $\log$ $g$ through ionization equilibrium (i.e. 
$\log$ $g$ is constrained by imposing that both neutral and ionized iron lines give the same abundance) to be 
systematically biased towards lower gravities. Thus, we prefer to rely on photometric gravities,
derived by using the Stefan-Boltzmann relation. We adopted
an absolute distance modulus of (m-M)$_{0}=14.4$ mag (from \citealt{harris}), bolometric corrections from
\cite{alonso99} and a mass of $0.75 \msol$. Such a mass has been derived from the best fit 
isochrone taken from the BaSTI dataset (\citealt{pie06}), with an age of 12 Gyr, Z=0.004 and $\alpha$-enhanced
chemical mixture (corresponding to [Fe/H]$=-1.01$ dex).
Uncertainties in $\log$ $g$ have been computed by taking into account the errors on T$_{{\rm eff}}$ (as described 
above), bolometric luminosity (due to all the photometric uncertainties) and mass (we assumed $\pm0.05 \msol$, that
corresponds to the range of masses that populate the entire RGB according to the best-fit isochrone).
The final uncertainty on $\log$ $g$ is about $0.05$ dex.

Finally, microturbulent velocities v$_{turb}$ have been derived spectroscopically, by requiring no trend
between the measured iron abundances and the line strengths. Typical uncertainties on this parameter
are about $0.1$ km$\,$s$^{-1}$.
The atmospheric parameters for each analysed target are shown in Table \ref{tab1}.

\subsection{Abundance measurements}

The adopted linelist has been compiled by selecting only transitions that are unblended at the 
temperatures, gravities, and metallicities of sampled stars.
Atomic data are from the lastest version of the Kurucz-Castelli database, improved for 
some specific cases with new and updated values.

For the elements in our linelist with single and unblended lines we estimated the chemical abundances 
from the measured equivalent width (EW), by using the package GALA \citep{gala}\footnote{GALA is freely 
distributed at the Cosmic-Lab project website, {\tt http://www.cosmic-lab.eu/gala/gala.php}}.
We run GALA keeping T$_{{\rm eff}}$ and $\log$ $g$ of the model fixed and allowing its metallicity to vary iteratively in order to 
match the iron abundance measured from EWs. All the model atmospheres have been computed by means 
of the ATLAS9 code (\citealt{atlas}), while EWs were measured by using DAOSPEC through the code 4DAO (see Section \ref{vrad}).
EW uncertainties are estimated by DAOSPEC as the standard deviation of the local flux residuals \citep[see][]{daospec}.  
All the lines with EW errors larger than 10\% were excluded from the analysis.
Solar reference abundances are taken from \cite{gs98}.

Only for the lines with hyperfine structure and/or isotopic splitting (Cu, Ba, Nd, Eu, La), abundances have been
obtained by individually comparing the observed spectral lines with a grid of synthetic spectra computed with 
SYNTHE (\citealt{sbordone04}), by running a $\chi^2$-minimization algorithm 
\citep[see the procedure described in][]{m12_2419}. In particular, synthetic spectra -- computed by assuming for each star the appropriate
atmospheric parameters derived as described in Section~\ref{atm} --  are convolved at the UVES resolution and finally re-sampled at the pixel 
size as the observed spectra.

\subsection{Abundance uncertainties}

The total uncertainties on abundance measurements has been derived by considering two main sources
of error: the internal error associated to the measurement procedure and the errors arising from
the uncertainties in the atmospheric parameters.

We defined as internal error the dispersion around the mean divided by the 
square root of the number of lines used to compute abundances.
Abundances of elements measured via spectral synthesis method
(namely Cu, La and Eu) come from the measurement of only one line. In this case, the corresponding
internal error has then been computed by means of Monte Carlo simulations. 
Briefly, poissonian noise is added to the best-fit synthetic spectrum in order to reproduce the observed SNR and then 
the procedure to derive the abundance is repeated.  The dispersion of the abundances measured from 1000 Monte Carlo
realizations has been adopted as the internal abundance uncertainty. Depending on the SNR around these lines
(ranging from $30$ to $60$), the typical internal errors obtained in this way range from $0.03$ dex up to $0.11$ dex.

In order to quantify the error coming from the uncertainties on the atmospheric parameters,
we repeated the chemical analysis by varying each parameter for the corresponding uncertainty (see Section \ref{atm}).

The total uncertainty on the [X/H] abundances has been computed by summing in quadrature this contribution and the intrinsic error described above. 
As discussed by in \citealt{mc95} this kind of uncertainty, when related to abundance 
ratios as [X/Y]=[X/H]-[Y/H], partially cancel out because lines of the same ionization stage tend to react in a similar way
to changes in the stellar parameters. Therefore, we followed their prescription to calculate our final [X/Fe] abundance uncertainties.
 
\section{RESULTS}\label{res}

In this Section the results of the chemical analysis are described for each elemental group. In particular,
elemental abundances as found for NGC~6362 (always shown as a red empty star symbol in Figures 3-8)
are directly compared to abundances taken from the literature for other GCs
(values from \citealt{carretta09} are shown as green filled circles\footnote{Ti abundances for several
GCs have been provided by E.Carretta, private communication}, while data from \citealt{apogee15}
are plotted as blue filled circles, if not stated otherwise) 
and Galactic field stars (grey dots, data from from \citealt{ful2000, gratton03, reddy03, reddy06}).
All the measured values and the corresponding uncertainties are listed in Tables \ref{tab2}, \ref{tab3} and \ref{tab4}, while 
the average abundance ratios are listed in Table \ref{tab5}.

\begin{itemize}
 \item {\it Iron abundances}.
 
By using the Maximum Likelihood (ML) algorithm described in \cite{m12_2419}, 
we found that the iron distribution for our spectroscopic sample is best described by a Gaussian function 
with mean $<$[FeI/H]$>=-1.07\pm0.01$ and a null dispersion of $\sigma_{{\rm[FeI/H]}}=0.00\pm0.01$. 
The derived Fe abundance agrees well with the value provided by \citet{muccia16}, i.e.  [FeI/H]$=-1.09\pm0.01$ dex.
A very similar value is obtained from the single ionized Fe~II lines that provide an average abundance of 
$<$[FeII/H]$>=-1.06\pm0.01$, with null dispersion. \\

\item {\it Light elements: Mg, Al}.

As already demonstrated by \cite{muccia16} for a larger sample of stars including the 11 targets studied here, 
NGC~6362 displays a broad [Na/Fe] distribution, with the presence of 
two (equally populated) stellar groups, peaked at [Na/Fe]$=+0.00$ dex and [Na/Fe]$=+0.33$ dex.
Another significant feature related to the existence of MPs in the form of light element abundance spreads is the presence of star-to-star variations in Mg and Al abundances. 
This feature arises as an anti-correlation in a few clusters,  whereas a large Al variation is coupled with a small or null Mg variation in most cases \citep{carretta09}.

\begin{figure}
\includegraphics[width=\columnwidth]{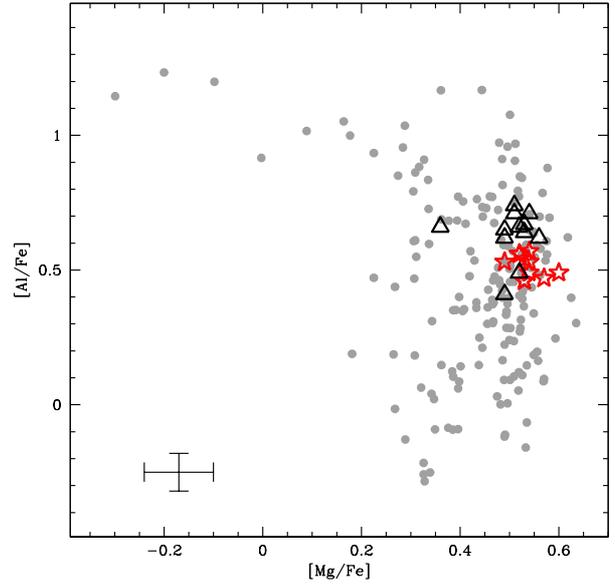}
\caption{\small [Al/Fe] abundance ratio as a function of [Mg/Fe] for the $11$ UVES targets of NGC~6362 analysed in this work (red stars), 
for 12 giants belonging to M~4 (black triangles), and for $17$ Galactic GCs (grey dots, data from \citealt{carretta09}). 
Typical errors are shown in the bottom-left region of the plot.}
\label{mgal}
\end{figure}

As shown in Figure \ref{mgal}, NGC~6362 does not display intrinsic variations in both Mg and Al. In fact NGC~6362 stars 
describe only a clump with very small dispersion compared to the whole extent of the anti-correlation covered by 
the UVES targets analysed in \cite{carretta09} in 17 GCs (grey dots).
According to a ML analysis, the mean abundances and intrinsic dispersions for these 
two elements are $<$[Mg/Fe]$>=+0.54\pm0.01$,  $\sigma_{{\rm[Mg/Fe]}}=0.00\pm0.01$ and 
$<$[Al/Fe]$>=+0.51\pm0.02$,  $\sigma_{{\rm[Al/Fe]}}=0.00\pm0.02$, thus confirming the lack for any abundance 
spread in both [Mg/Fe] and [Al/Fe].

 \begin{figure}
\includegraphics[width=\columnwidth]{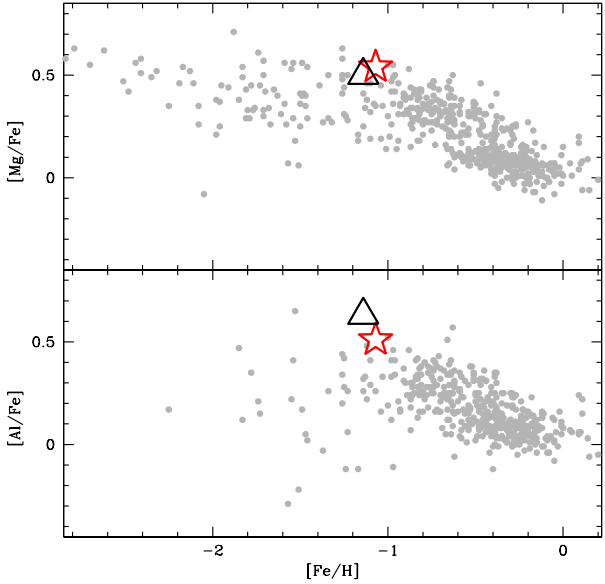}
\caption{\small {\it Upper panel:} [Mg/Fe] vs [Fe/H] trends for NGC~6362 (red star) and M~4 (black triangle), compared to a sample of field stars (grey dots) . 
{\it Lower panel:} same comparison for [Al/Fe].}
\label{mgal_gc}
\end{figure}

In Figure \ref{mgal_gc} the average Mg and Al abundances of NGC~6362 stars are compared to those of
Galactic field stars\footnote{We decided not to display other Galactic GCs in the plot, as for those showing intrinsic spread
in Mg and Al abundance an average value is not meaningful.}.
Although NGC~6362 stars appear to be quite rich in both Mg and Al content with respect 
to stars at similar same metallicity (upper and lower panel of Figure~\ref{mgal_gc}, respectively), their Mg and Al  
abundances are still in agreement with the trends observed for field stars.\\

\item {\it $\alpha$-elements: Si, Ca, Ti}.

We adopt the $\alpha$-elements produced by explosive nucleosynthesis (Si, Ca and Ti) as tracers of the total 
$\alpha$-element abundance for the cluster. Mg (that is produced by hydrostatic nucleosynthesis) 
is excluded from this discussion because self-enrichment 
processes can, in principle, alter its initial abundance (even if in NGC~6362 no evidence of a Mg spread is found).

\begin{figure}
\includegraphics[width=\columnwidth]{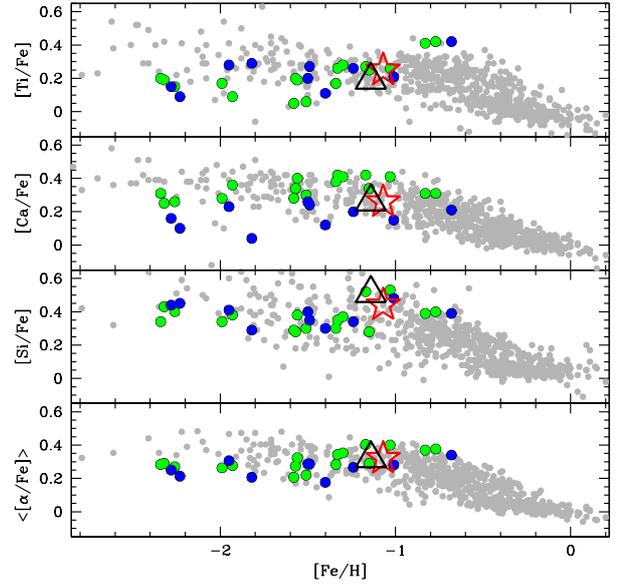}
\caption{\small {\it Upper three panels:} [Si/Fe], [Ca/Fe], [Ti/Fe] vs. [Fe/H] trends for NGC~6362, field stars  (same symbols of Figure \ref{mgal_gc})
and Galactic GCs (green circles are taken from \citealt{carretta09}, blue circles from \citealt{apogee15}). 
{\it Lower panel:} same comparison on the [$\alpha$/Fe] vs. [Fe/H] trend. NGC~6362 is in very good agreement 
with what observed for the other GCs.}
\label{sicati}
\end{figure}

When analysed separately, Si is the element with the highest average enhancement, having $<$[Si/Fe]$>=+0.45\pm0.03$, 
while Ca and Ti show lower values that are very similar each other, $<$[Ca/Fe]$>=+0.26\pm0.02$ and 
$<$[Ti/Fe]$>=+0.24\pm0.04$, respectively.
According to the ML analysis, none of the three elements show any hints of intrinsic dispersion. 
Figure \ref{sicati} shows the comparison among NGC~6362, and the same objects as those shown in Fig.\ref{mgal_gc}.
Also in terms of Si, Ca, and Ti content, NGC~6362 does not behave differently from what commonly is observed 
for the other populations.\\

\item {\it Iron-peak elements: Sc, V, Mn, Cr, Co, Ni}.

We measured abundances for six iron-peak elements, namely Sc, V, Mn, Cr, Co and Ni.

For Cr and Ni the abundances have been derived from the EW measurement, and both 
the abundances of these elements turn out to be scaled-solar,
with average values of $<$[Cr/Fe]$>=-0.05\pm0.04$ and $<$[Ni/Fe]$>=-0.02\pm0.01$. 

For odd-Z elements like Sc and Mn, we used the spectral synthesis method since their lines suffer for hyperfine splitting.    
In this case we found a super-solar [Sc/Fe] abundance ratio, with an average $<$[ScII/Fe]$>=+0.18\pm0.01$, while 
we found a significant deficiency in the Mn content of the cluster ($<$[Mn/Fe]$>=-0.33\pm0.02$).

Finally, the presence of both isolated lines and lines split for hyperfine structure for V and Co within the UVES wavelength range 
allowed us to check for possible systematic effects arising from the different method of analysis used. For both the elements we found that the
abundances measured with the EW method and those coming from the spectral synthesis agree very well, being coincident within
a $1-\sigma$ uncertainty. In particular we obtained slightly super-solar abundance ratios for both V ($<$[V/Fe]$_{EW}>=+0.07\pm0.04$ and 
$<$[V/Fe]$_{synthesis}>=+0.10\pm0.05$, respectively) and Co ($<$[Co/Fe]$_{EW}>=+0.12\pm0.02$ and 
$<$[Co/Fe]$_{synthesis}>=+0.09\pm0.02$). Therefore we can safely claim that no systematic uncertainties are introduced by the use 
of two different abundance measurement methods.

\begin{figure}
\includegraphics[width=\columnwidth]{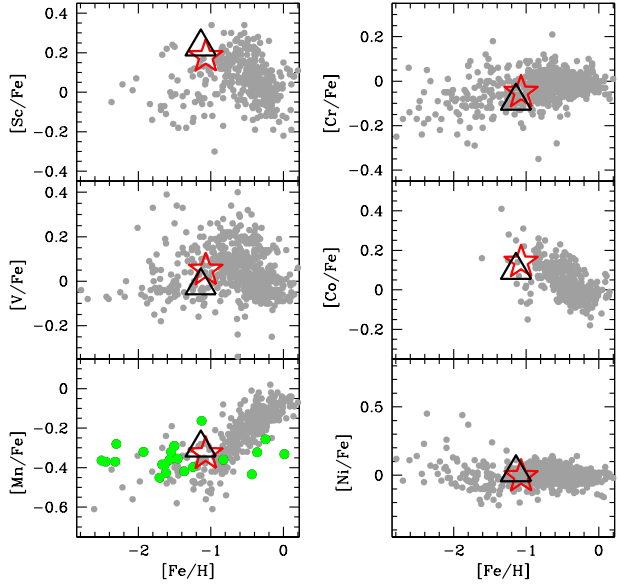}
\caption{\small Comparison among the trends of the iron-peak element abundances vs. [Fe/H] as measured for NGC~6362, M~4 
and field stars. Green circles indicate the [Mn/Fe] average abundances derived by \citep{sobeck06} for 21 Galactic GCs.}
\label{ironpeaks}
\end{figure}

The iron-peak elemental abundances for NGC~6362 are compared to the typical values found for field stars in Fig~\ref{ironpeaks}.\\

\item {\it Copper, s- and r- elements}.

The abundances for these elements have all been measured by means of spectral synthesis.

For Cu, the only available transition is that at 5105 \AA, since the other Cu optical line 
(at 5782 \AA) falls in the gap between the two UVES chips.
NGC~6362 behaves similarly to the other (few) GCs in the Galaxy for which Cu measurements exist
(see \citealt{cunha02, simmerer}), with an average $<$[Cu/Fe]$>=-0.18\pm0.03$. This is also in good agreement 
with measurements for field stars at similar [Fe/H] (see the upper panel of Figure \ref{copper}). 

We determined the abundance of the slow neutron-capture elements Ba, Nd and La.
Regarding the abundance of Ba, we measured $<$[BaII/Fe]$>=+0.56\pm0.01$.
By using only the La line at 6390.5 \AA\ \citep{laline} we derived
$<$[LaII/Fe]$>=+0.36\pm0.02$, while for Nd we found $<$[Nd/Fe]$>=+0.37\pm0.02$. 

Finally we estimated the abundance of the rapid neutron-capture element Eu by using the
transition at 6645.1 \AA, finding $<$[Eu/Fe]$>=+0.43\pm0.01$. NGC~6362 matches well
the typical behaviour of [Eu/Fe] measured for the other GCs in the Galaxy (see the bottom panel of Figure \ref{copper},
where data for Eu abundances in 14 GCs have been taken from 
\citealt{sneden97, sneden04, ivans99, ivans01, carretta04, carretta06, ramirez02, lee02, james04, yong05, marino15}).

 \begin{figure}
\includegraphics[width=\columnwidth]{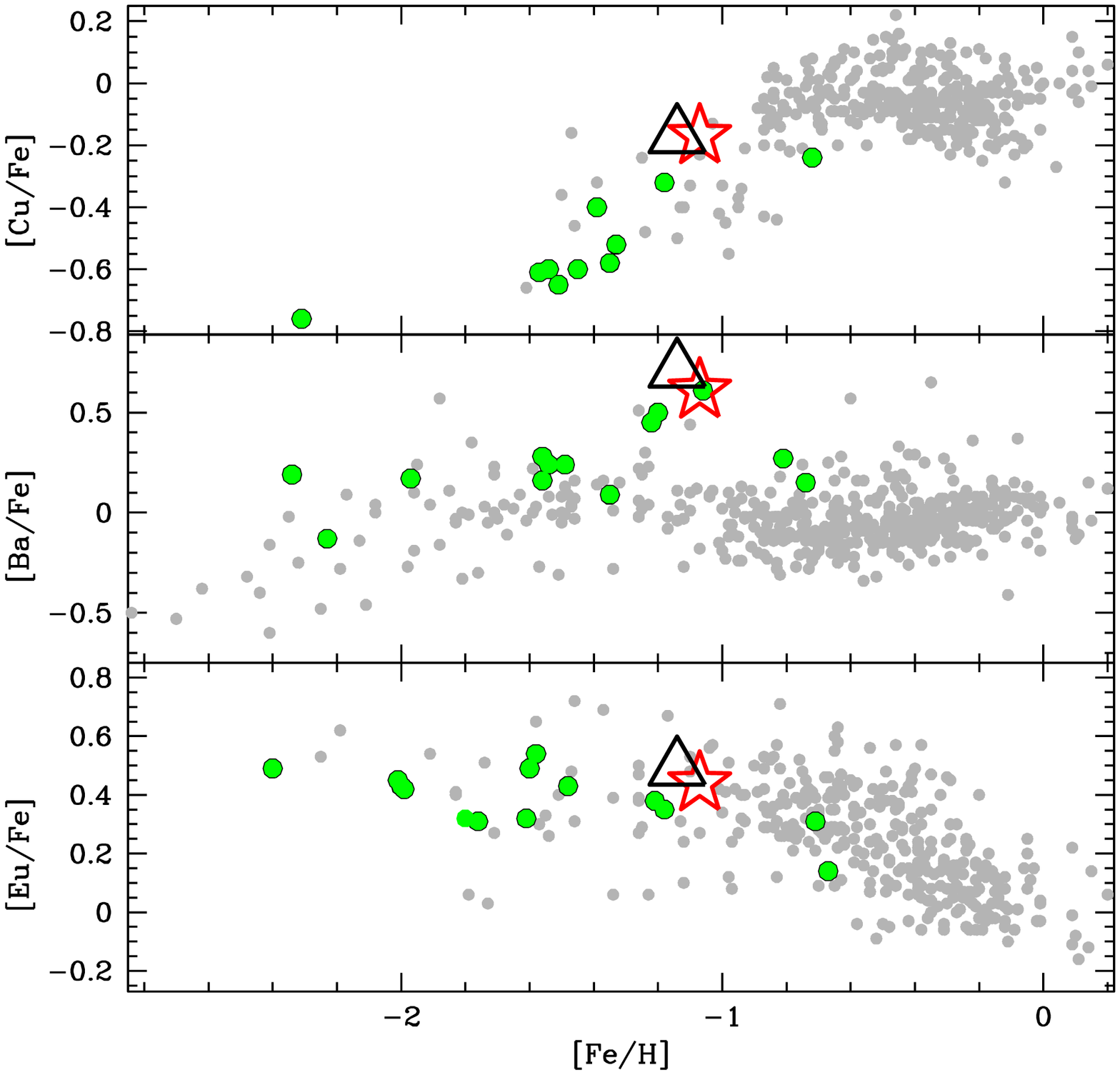}
\caption{\small {\it Upper panel:} [Cu/Fe] vs [Fe/H] trends for a sample of Galactic GCs (green circles) taken from \citep{simmerer}, a sample of thick disk stars
taken from \citep{reddy03, reddy06} and the cluster under analysis. {\it Central panel:} observed Ba trends for the same populations
(data for the 14 plotted GCs have been taken from \citealt{dorazi}). {\it Lower panel:} same comparison but for Eu abundances. }
\label{copper}
\end{figure}

\end{itemize}

\subsection{Analysis of M~4 as a reference}

In the previous paragraphs we compared the elemental abundances of NGC~6362 with those of field stars and other
GCs available in the literature. However, such a comparison could be prone to systematic effects due for 
example to different used instruments or adopted linelists, atomic parameters, model atmospheres etc. 
To provide a further comparison free from all of these systematics, we decided to repeat our analysis on 
another GC with mass (see \citealt{mcvan05}) and metallicity very similar to those of NGC~6362: M~4.
This differential comparison between two clusters with similar properties and 
analysed with the same procedure will allow us to accurately highlight any intrinsic difference among 
their chemical composition.
Following the same approach described by \cite{muccia13}, we thus re-analysed 12 RGB stars belonging to 
M~4 and observed with FLAMES-UVES Red Arm 580 (ID 073.0211, PI:Carretta). 
B and V magnitudes for M~4 have been obtained from the analysis of a dataset of WFI\@2.2m images 
(program 68.D-0265(A), PI: Ortolani), and have been corrected for differential reddening. 
The distance modulus ($(m-M)_{0}=11.78$ mag) and colour excess (E(B-V)$=0.32$) mag have been taken from 
\cite{bedin09}. Atmospheric parameters for the M4 targets have been obtained following the same 
approach used for the NGC~6362 stars.
The results of such an investigation are shown in Figures 3-8, where abundance ratios for M~4 are always shown
as black triangles. The average values are listed in Tab.\ref{tab5}, where the one-to-one comparison between
the NGC~6362 and M~4 is displayed.

As a first step, we checked that the abundance ratios we obtained as output of our analysis were in agreement with
previous results available in the literature.
The iron abundance we measured for M~4 is $<$[Fe/H]$>=-1.14\pm0.01$. This value is in good 
agreement with previous metallicity determinations obtained from giant stars 
(see \citealt{carretta09, muccia11, villanova11, monaco12}. 
Mg and Al abundance ratios are shown in Fig.\ref{mgal} as black triangles. Like NGC~6362 stars, they describe
a compact clump as well, with no dispersion in Mg and only a small hint of dispersion in Al,
confirmed by the ML analysis which found $\sigma_{{\rm[Al/Fe]}}=0.05\pm0.03$ (a small Al dispersion
for the cluster has also been found by \citealt{marino08}). Their average values are shown in Fig.~\ref{mgal_gc},
and agree well with the results in \cite{marino08} and \cite{ivans99}, as $\alpha$- and iron-peak elemental abundances do.
It is worth noticing that for M~4 \cite{simmerer} found $<$[Cu/Fe]$>=-0.32$ dex
by using both the Cu lines previously described. These authors used the same solar values and oscillator
strength for the line used by us, and they obtained agreement within $0.1$ dex between measurements coming from the two
lines individually. Therefore their estimate is compatible with ours to within $\sim1\sigma$.
Finally, our measurement of $<$[Ba/Fe]$>=+0.71\pm0.04$ for M~4 is $\sim0.3$ dex larger 
than that found in \cite{marino08}. However, these authors used a solar value A(Ba)$\odot=2.45$ instead of our A(Ba)$\odot=2.13$.
Therefore such a discrepancy cancels out when the different adopted solar values are taken into account.

\begin{figure}
\includegraphics[width=\columnwidth]{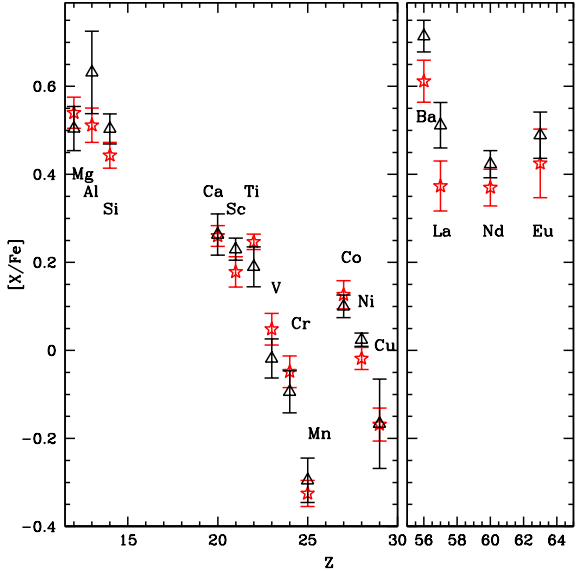}
\caption{\small Comparison among the elemental abundances measured for NGC~6362 (red star symbols) and M~4 
(black triangles).}
\label{valmedi}
\end{figure}

At this point we performed the direct, one-to-one comparison between the chemistry of the two clusters, shown in Fig.~\ref{valmedi}.
In general, we found that the agreement between the elemental abundances of NGC~6362 and M~4 is remarkable.
In fact, all the elemental abundance ratio match within a $1\sigma$ uncertainty, with the only exception of [BaII/Fe] and [LaII/Fe],
for which the agreement is only within a $2\sigma$ uncertainty.

\section{DISCUSSION AND CONCLUSIONS}\label{discuss}

We measured abundance ratios for different elemental groups in the Galactic GC NGC~6362, 
which has never been investigated before using high-resolution spectroscopy, and we compared it 
with other GCs in the Milky Way.

\begin{figure}
\includegraphics[width=\columnwidth]{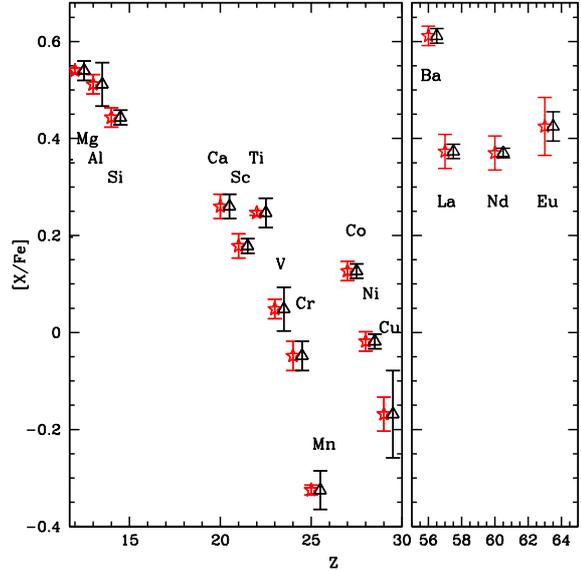}
\caption{\small Comparison among the elemental IQR for NGC~6362 (red bars) and M~4 (black bars). For sake of comparison, all the pars are plotted
at the same y-axis value (that measured for NGC~6362).}
\label{interq}
\end{figure}

With the aim of anchoring our findings to a solid touchstone, we also repeated the same chemical analysis 
(thus erasing any systematics) on a sample of RGB stars in M~4,
a cluster that shares with NGC~6362 very similar mass and metallicity. 
Fig.~\ref{valmedi} and~\ref{interq} show the comparison of the measured abundance ratios 
in NGC~6362 (red star symbols) and M~4 (black triangles), using the average values and the 
interquartile-range (IQR), respectively.

\begin{itemize}
 \item {\it Light elements.} 
 
 \cite{muccia16} demonstrated that NGC~6362 hosts two distinct populations
 with different Na abundances. While Na star-to-star variations have been observed in most GCs, Mg and Al
 intrinsic spreads are not as universal. In particular, according to both observations (see e.g. \citealt{carretta09}) 
 and theoretical predictions (\citealt{ventura14}), the Al production in GCs depends on two main parameters: 
 ($i$) metallicity, since a high metallicity causes lower Al-yields from polluters\footnote{this is true
 for AGB polluters, while no detailed Al-yields predictions are currently available for Fast Massive Rotating 
 Stars (\citealt{decressin07}) or supermassive stars (\citealt{denissenkov})} (\citealt{ventura09, karakas10, 
 oconnell11}) and ($ii$) cluster mass, since (under the hypothesis of the same initial mass function) a 
 more massive cluster would have more polluters and would retain more polluted gas within its potential well.
 However, several exceptions to these trends exist, such as 47 Tucanae (which shows a small Al spread despite 
 being quite massive, see \citealt{cordero14}) or M71 (which is an outlier in the Carretta relation, see 
 \citealt{carretta09, cordero15}). Therefore, the Al abundances measured in this work for NGC~6362 and their 
 comparison to M~4, are of particular interest.
 
 As already shown is Sect.~\ref{res}, Mg and Al internal variations in NGC~6362 are compatible with being null (see also 
 Fig.\ref{interq}), while M~4 stars show a possible small dispersion in Al content. 
 The corresponding IQRs are extremely small as well, being of few hundredths 
 of dex for Mg and only slightly larger for Al (IQR$^{NGC~6362}_{Al}=0.04$ dex, IQR$^{M~4}_{Al}=0.09$ dex).
 Therefore, we can conclude that in terms of light element abundance, NGC~6362 behaves as the other GCs with 
 similar mass and metallicity, both according to the differential analysis with respect to M~4 and to 
 the observational-based prediction of \cite{carretta09}.
 
 \item {\it $\alpha$- and iron-peak elements.} 
 
 NGC~6362 is characterized by enhanced abundances for all the measured 
 $\alpha$-elements, with an average value of $<$[$\alpha$/Fe]$>=+0.32$. This value matches well those measured 
 in M~4 and in the other Galactic GCs \citep[see e.g.][]{pritzl05,carretta10,apogee15}.
 This finding confirms that NGC~6362 formed from gas enriched by core-collapse supernovae in a similar way to other Galactic 
 GCs (excluding some GCs likely accreted by extragalactic galaxies and characterized by lower [$\alpha$/Fe] ratios). 
 In particular, NGC~6362 and M~4 show very similar $\alpha$ abundances, suggesting that they share a similar 
 chemical enrichment by core-collapse supernovae. 
 Also for the iron-peak elements, the analysis shows that NGC~6362 follows the same abundance patterns of the other 
 GCs and no remarkable (element-by-element) difference is found between NGC~6362 and M~4.
 
 \item {\it s- and r-process elements} 
  
 The cluster has a [Ba/Fe] abundance ratio compatible with those of GCs of similar 
 metallicity, like M4, NGC~288 and NGC~6171, and, in general, agrees well with the run 
 of [s/Fe] abundance ratios with [Fe/H] \citep[see Fig.1 in][]{dorazi}. In fact, metal-poor GCs show roughly solar [Ba/Fe] ratios, 
 as the production of Ba at those metallicities is dominated by r-processes. 
 On the other hand [Ba/Eu] increases with increasing the metallicity, because of the higher efficiency of 
 the s-process. In the case of NGC~6362, [Ba/Eu] reaches $0.18$ dex, compatible 
 with the value measured in M~4 ([Ba/Eu]$=0.22$ dex) and the same is observed also for [La/Eu] and [Nd/Eu]. 
 Even if NGC~6362 has a [Ba/Fe] compatible 
 with that of M~4 at a level of 2$\sigma$ (at variance with the other abundance ratios that are compatible 
 within 1$\sigma$), the two clusters show the same relative efficiency of s- to r-process.
 Therefore, the general enhancement of s-process element abundances measured in NGC~6362 suggests that the first generation of cluster stars 
 $i$) formed from gas already enriched by low-mass ($\sim1-4M_{\odot}$) AGB stars, that are the main producers of s-process elements,
 the so-called main-component \citep[see e.g][]{busso99,travaglio04}, and $ii$) did not introduce any spread the s-process abundance
 of second generation stars.

\end{itemize}

Summing all up, the differential comparison between NGC~6362 and M~4 revealed that
all the elemental abundances measured for the two clusters match within 1-$\sigma$ 
(with the marginal exception of [Ba/Fe]). 
None of the elements analysed in this work show any internal spread in NGC~6362.
It is worth noting that NGC~6362 and M~4 also display the same extent of Na variation 
\citep[see][]{muccia16}.

According to the compilation of GC masses in \cite{mcvan05}, and of GC metallicities 
in \cite{harris}, the only clusters with properties similar to NGC~6362 and M~4 (and not associated 
with the Sagittarius dwarf spheroidal galaxy, see \citealt{law10}) which have been chemically investigated 
in detail are NGC~288 and NGC~6171. They both belong to the sample of clusters studied by \cite{carretta09}, 
and as such they are plotted in Fig.\ref{sicati}, in the same metallicity range as that of NGC6362. 
Their chemical composition is consistent with that of the two clusters analysed in this work.
Therefore, we conclude that NGC~6362 is a regular GC that shows the chemical composition 
representative of the Milky Way GCs with similar mass and metallicity, with the signatures 
of chemical enrichment by core-collapse supernovae and AGB stars.

\section*{Acknowledgements}

We thank the anonymous referee for her/his comments and suggestions 
which improved the presentation of our results.
DM and GF has been supported by the FIRB 2013 (MIUR grant RBFR13J716).
SC acknowledges financial support from PRIN - INAF 2014 (PI: S.Cassisi).
  This research is part of the project COSMIC-LAB
  (web site: http://www.cosmic-lab.eu) funded by the European Research
  Council (under contract ERC-2010-AdG-267675). M.B. acknowledges financial 
  support from PRIN MIUR 2010-2011 project “The Chemical and Dynamical 
  Evolution of the Milky Way and Local Group Galaxies”, prot. 2010LY5N2T.

\begin{landscape}
\begin{deluxetable}{cccccccccc}
\tablewidth{0pc}
\tablecolumns{10}
\tiny
\tablecaption{NGC~6362 UVES targets analysed in this work.}
\tablehead{\colhead{ID} & \colhead{RA} & \colhead{Dec} & \colhead{B} & \colhead{V} & \colhead{T$_{eff}$} & \colhead{log~$g$} & \colhead{v$_{turb}$} & \colhead{v$_{rad}$} & \colhead{$\sigma_{vrad}$}\\
 & (degrees) & (degrees) & \colhead{(mag)} & \colhead{(mag)} & \colhead{(K)} & \colhead{ } & \colhead{( km$\,$s$^{-1}$)} & \colhead{( km$\,$s$^{-1}$)} & \colhead{ (km$\,$s$^{-1}$)} }
\startdata
 & & \\
        601063  &    262.8751296   &    -67.0970861  & 15.023 &  13.789 & 4308 & 1.33 & 1.5  & -14.39 & 0.06 \\
        602339  &    262.8462133   &    -67.0100153  & 14.919 &  13.674 & 4292 & 1.27 & 1.5  & -14.23 & 0.12 \\
        709358  &    263.0091350   &    -67.0564120  & 14.744 &  13.393 & 4147 & 1.05 & 1.5  & -15.24 & 0.07 \\
        710376  &    262.9964883   &    -67.0425252  & 15.099 &  13.916 & 4382 & 1.43 & 1.4  & -17.39 & 0.13 \\
        711565  &    262.9825550   &    -67.0373852  & 15.058 &  13.845 & 4338 & 1.37 & 1.5  & -10.75 & 0.07 \\
        716150  &    262.9265237   &    -67.0608503  & 15.086 &  13.902 & 4381 & 1.42 & 1.5  & -16.01 & 0.06 \\ 
        601269  &    262.8706968   &    -67.1041887  & 14.706 &  13.355 & 4147 & 1.04 & 1.6  & -13.42 & 0.09 \\
        604027  &    262.7885538   &    -67.1157141  & 14.848 &  13.579 & 4259 & 1.21 & 1.5  & -14.98 & 0.08 \\
        703323  &    263.1077119   &    -67.0442269  & 14.658 &  13.282 & 4114 & 0.98 & 1.5  & -17.33 & 0.09 \\
        714494  &    262.9480400   &    -67.0372456  & 14.644 &  13.296 & 4151 & 1.02 & 1.5  & -14.02 & 0.07 \\
        716237  &    262.9253519   &    -67.0381319  & 14.610 &  13.185 & 4051 & 0.90 & 1.5  & -17.93 & 0.11 \\
\enddata
\tablecomments{\small Identification number, coordinates, B and V magnitudes, atmospheric parameters, heliocentric 
radial velocities and their uncertainties for the 11 UVES targets of NGC~6362 analysed in this work. The photometric parameters
have been taken from the catalogue of \cite{dalex14}}.
\label{tab1}
\end{deluxetable}
\end{landscape}

\begin{landscape}
\tiny
\begin{deluxetable}{ccccccc}
\tablewidth{0pc}
\tablecolumns{7}
\tablecaption{Elemental abundances for the analysed targets: [Fe/H], light- and $\alpha$-elements.}
\tablehead{\colhead{ID}  & \colhead{[Fe/H]}  &  \colhead{[Mg/Fe]} & \colhead{[Al/Fe]} &  \colhead{[Si/Fe]} & \colhead{[Ca/Fe]} & \colhead{[Ti/Fe]} }  
\startdata
 & & \\
        601063 & $-1.06 \pm 0.05$ & $0.53 \pm 0.05$ & $0.46 \pm 0.06$ & $0.45 \pm 0.12$ & $0.29 \pm 0.08$ & $0.24 \pm 0.14$ \\
        602339 & $-1.08 \pm 0.05$ & $0.54 \pm 0.06$ & $0.49 \pm 0.06$ & $0.43 \pm 0.13$ & $0.24 \pm 0.08$ & $0.25 \pm 0.14$ \\
        709358 & $-1.07 \pm 0.02$ & $0.54 \pm 0.04$ & $0.53 \pm 0.07$ & $0.46 \pm 0.10$ & $0.25 \pm 0.08$ & $0.28 \pm 0.14$ \\
        710376 & $-1.04 \pm 0.06$ & $0.49 \pm 0.08$ & $0.53 \pm 0.07$ & $0.42 \pm 0.13$ & $0.29 \pm 0.08$ & $0.24 \pm 0.13$ \\
        711565 & $-1.11 \pm 0.06$ & $0.60 \pm 0.09$ & $0.49 \pm 0.06$ & $0.49 \pm 0.14$ & $0.25 \pm 0.09$ & $0.24 \pm 0.13$ \\
        716150 & $-1.06 \pm 0.06$ & $0.54 \pm 0.08$ & $0.57 \pm 0.07$ & $0.41 \pm 0.13$ & $0.24 \pm 0.09$ & $0.23 \pm 0.13$ \\ 
        601269 & $-1.07 \pm 0.04$ & $0.57 \pm 0.04$ & $0.47 \pm 0.07$ & $0.43 \pm 0.13$ & $0.28 \pm 0.08$ & $0.26 \pm 0.13$ \\
        604027 & $-1.07 \pm 0.05$ & $0.52 \pm 0.05$ & $0.56 \pm 0.08$ & $0.46 \pm 0.12$ & $0.29 \pm 0.09$ & $0.22 \pm 0.14$ \\ 
        703323 & $-1.10 \pm 0.04$ & $0.52 \pm 0.04$ & $0.56 \pm 0.07$ & $0.49 \pm 0.14$ & $0.24 \pm 0.08$ & $0.21 \pm 0.13$ \\ 
        714494 & $-1.08 \pm 0.04$ & $0.53 \pm 0.05$ & $0.54 \pm 0.07$ & $0.45 \pm 0.12$ & $0.28 \pm 0.08$ & $0.26 \pm 0.14$ \\ 
        716237 & $-1.03 \pm 0.05$ & $0.53 \pm 0.04$ & $0.46 \pm 0.08$ & $0.42 \pm 0.15$ & $0.26 \pm 0.09$ & $0.23 \pm 0.12$\\ 
\enddata
\tablecomments{\small Identification number, elemental abundances and related uncertainties for the same 11 targets described in Tab.\ref{tab1}. This Table continues
in Tab.\ref{tab3} and Tab.\ref{tab4} for other groups of elements.}
\label{tab2}
\end{deluxetable}
\end{landscape}

\begin{landscape}
\tiny
\begin{deluxetable}{cccccccc}
\tablewidth{0pc}
\tablecolumns{8}
\tablecaption{Elemental abundance for the analysed targets: iron-peak elements}
\tablehead{\colhead{ID} & \colhead{[Sc/Fe]} & \colhead{[V/Fe]} & \colhead{[Cr/Fe]} &  \colhead{[Mn/Fe]} & \colhead{[Co/Fe]} & \colhead{[Ni/Fe]} & \colhead{[Cu/Fe]} }
\startdata
 & & \\
        601063 & $0.14 \pm 0.06$ & $0.03 \pm 0.15$ & $-0.09 \pm 0.14$ & $-0.34 \pm 0.09$ & $0.13 \pm 0.10$ & $ 0.00 \pm 0.06$ & $-0.14 \pm 0.13$ \\
        602339 & $0.20 \pm 0.05$ & $0.07 \pm 0.15$ & $-0.05 \pm 0.15$ & $-0.32 \pm 0.10$ & $0.14 \pm 0.06$ & $-0.04 \pm 0.07$ & $-0.15 \pm 0.13$ \\
        709358 & $0.19 \pm 0.03$ & $0.11 \pm 0.14$ & $-0.03 \pm 0.15$ & $-0.32 \pm 0.09$ & $0.17 \pm 0.05$ & $ 0.00 \pm 0.04$ & $-0.21 \pm 0.12$ \\
        710376 & $0.16 \pm 0.07$ & $0.04 \pm 0.15$ & $ 0.00 \pm 0.14$ & $-0.40 \pm 0.10$ & $0.08 \pm 0.08$ & $-0.03 \pm 0.08$ & $-0.12 \pm 0.13$ \\
        711565 & $0.23 \pm 0.07$ & $0.01 \pm 0.13$ & $-0.03 \pm 0.15$ & $-0.34 \pm 0.10$ & $0.14 \pm 0.10$ & $ 0.01 \pm 0.07$ & $-0.21 \pm 0.13$ \\
        716150 & $0.15 \pm 0.07$ & $0.03 \pm 0.14$ & $-0.09 \pm 0.15$ & $-0.33 \pm 0.10$ & $0.10 \pm 0.07$ & $-0.05 \pm 0.07$ & $-0.18 \pm 0.13$ \\ 
        601269 & $0.21 \pm 0.04$ & $0.11 \pm 0.13$ & $-0.02 \pm 0.14$ & $-0.29 \pm 0.09$ & $0.13 \pm 0.05$ & $-0.02 \pm 0.06$ & $-0.20 \pm 0.12$  \\
        604027 & $0.17 \pm 0.06$ & $0.03 \pm 0.14$ & $-0.02 \pm 0.15$ & $-0.34 \pm 0.10$ & $0.09 \pm 0.05$ & $-0.03 \pm 0.05$ & $-0.13 \pm 0.13$  \\ 
        703323 & $0.20 \pm 0.05$ & $0.08 \pm 0.13$ & $-0.03 \pm 0.13$ & $-0.34 \pm 0.10$ & $0.13 \pm 0.05$ & $-0.03 \pm 0.06$ & $-0.19 \pm 0.12$  \\ 
        714494 & $0.18 \pm 0.04$ & $0.10 \pm 0.13$ & $-0.01 \pm 0.14$ & $-0.33 \pm 0.10$ & $0.12 \pm 0.06$ & $-0.04 \pm 0.06$ & $-0.16 \pm 0.12$  \\ 
        716237 & $0.12 \pm 0.05$ & $0.18 \pm 0.15$ & $-0.02 \pm 0.15$ & $-0.34 \pm 0.09$ & $0.15 \pm 0.06$ & $-0.03 \pm 0.06$ & $-0.29 \pm 0.13$  \\ 
\enddata
\label{tab3}
\end{deluxetable}
\end{landscape}

\begin{landscape}
\tiny
\begin{deluxetable}{ccccc}
\tablewidth{0pc}
\tablecolumns{5}
\tablecaption{Elemental abundance for the analysed targets: s- and r- elements}
\tablehead{\colhead{ID} & \colhead{[Ba/Fe]} & \colhead{[La/Fe]} &  \colhead{[Nd/Fe]} &  \colhead{[Eu/Fe]} }
\startdata
        601063  & $0.63 \pm 0.13$ & $0.45 \pm 0.07$ & $0.26 \pm 0.08$ & $0.49 \pm 0.06$\\
        602339  & $0.60 \pm 0.11$ & $0.35 \pm 0.07$ & $0.25 \pm 0.07$ & $0.37 \pm 0.06$\\
        709358  & $0.64 \pm 0.12$ & $0.35 \pm 0.04$ & $0.22 \pm 0.04$ & $0.48 \pm 0.04$\\
        710376  & $0.53 \pm 0.13$ & $0.29 \pm 0.08$ & $0.18 \pm 0.08$ & $0.32 \pm 0.07$\\
        711565  & $0.67 \pm 0.12$ & $0.42 \pm 0.09$ & $0.36 \pm 0.08$ & $0.51 \pm 0.07$\\
        716150  & $0.60 \pm 0.12$ & $0.38 \pm 0.08$ & $0.24 \pm 0.09$ & $0.38 \pm 0.07$\\ 
        601269  & $0.66 \pm 0.11$ & $0.41 \pm 0.07$ & $0.29 \pm 0.07$ & $0.50 \pm 0.06$ \\
        604027  & $0.60 \pm 0.13$ & $0.35 \pm 0.08$ & $0.21 \pm 0.08$ & $0.43 \pm 0.05$ \\ 
        703323  & $0.63 \pm 0.12$ & $0.38 \pm 0.08$ & $0.24 \pm 0.09$ & $0.44 \pm 0.07$ \\ 
        714494  & $0.61 \pm 0.11$ & $0.37 \pm 0.07$ & $0.24 \pm 0.08$ & $0.43 \pm 0.05$ \\ 
        716237  & $0.54 \pm 0.02$ & $0.26 \pm 0.08$ & $0.15 \pm 0.08$ & $0.41 \pm 0.06$ \\ 
\enddata
\label{tab4}
\end{deluxetable}
\end{landscape}

\begin{landscape}
\begin{deluxetable}{ccc}
\tablewidth{0pc}
\tablecolumns{3}
\tablecaption{Comparison between NGC~6362 and M~4}
\tablehead{\colhead{Element} & \colhead{NGC~6362} & \colhead{M~4}}
\startdata
        $<$[Fe/H]$>$  & $-1.07 \pm 0.01$ & $-1.14 \pm 0.01$\\
        $<$[Mg/Fe]$>$ & $0.54  \pm 0.01$ & $0.50  \pm 0.02$\\
        $<$[Al/Fe]$>$ & $0.51  \pm 0.02$ & $0.63  \pm 0.02$\\
        $<$[Si/Fe]$>$ & $0.45  \pm 0.03$ & $0.50  \pm 0.02$\\
        $<$[Ca/Fe]$>$ & $0.26  \pm 0.02$ & $0.26  \pm 0.02$\\
        $<$[Sc/Fe]$>$ & $0.18  \pm 0.02$ & $0.23  \pm 0.01$\\
        $<$[Ti/Fe]$>$ & $0.24  \pm 0.04$ & $0.19  \pm 0.03$\\ 
        $<$[V/Fe]$>$  & $0.07  \pm 0.04$ & $-0.02 \pm 0.03$\\
        $<$[Cr/Fe]$>$ & $-0.05 \pm 0.04$ & $-0.09 \pm 0.03$\\ 
        $<$[Mn/Fe]$>$ & $-0.33 \pm 0.02$ & $-0.30 \pm 0.04$\\ 
        $<$[Co/Fe]$>$ & $0.12  \pm 0.02$ & $0.10  \pm 0.02$\\ 
        $<$[Ni/Fe]$>$ & $-0.02 \pm 0.01$ & $0.02  \pm 0.02$\\
        $<$[Cu/Fe]$>$ & $-0.18 \pm 0.03$ & $-0.17 \pm 0.03$\\
        $<$[Ba/Fe]$>$ & $0.61  \pm 0.01$ & $0.71  \pm 0.04$\\
        $<$[La/Fe]$>$ & $0.36  \pm 0.02$ & $0.51  \pm 0.02$\\
        $<$[Nd/Fe]$>$ & $0.37  \pm 0.02$ & $0.42  \pm 0.02$\\
        $<$[Eu/Fe]$>$ & $0.43  \pm 0.01$ & $0.49  \pm 0.01$\\
\enddata
\tablecomments{\small Average elemental abundances and related uncertainties as found for the two samples of giants in NGC~6362 and M~4.}
\label{tab5}
\end{deluxetable}
\end{landscape}












\bsp	
\label{lastpage}
\end{document}